# Design of a 3000-pixel transition-edge sensor x-ray spectrometer for microcircuit tomography

Paul Szypryt, Douglas A. Bennett, William J. Boone, Amber L. Dagel, Gabriella Dalton, W. Bertrand Doriese, Joseph W. Fowler, Edward J. Garboczi, Johnathon D. Gard, Gene C. Hilton, Jozsef Imrek, Edward S. Jimenez, Vincent Y. Kotsubo, Kurt Larson, Zachary H. Levine, John A. B. Mates, Daniel McArthur, Kelsey M. Morgan, Nathan Nakamura, Galen C. O'Neil, Nathan J. Ortiz, Christine G. Pappas, Carl D. Reintsema, Daniel R. Schmidt, Daniel S. Swetz, Kyle R. Thompson, Joel N. Ullom, Christopher Walker, Joel C. Weber, Abigail L. Wessels, and Jason W. Wheeler

*Abstract*—Feature sizes in integrated circuits have decreased substantially over time, and it has become increasingly difficult to three-dimensionally image these complex circuits after fabrication. This can be important for process development, defect analysis, and detection of unexpected structures in externally sourced chips, among other applications. Here, we report on a non-destructive, tabletop approach that addresses this imaging problem through x-ray tomography, which we uniquely realize with an instrument that combines a scanning electron microscope (SEM) with a transition-edge sensor (TES) x-ray spectrometer. Our approach uses the highly focused SEM electron beam to generate a small x-ray generation region in a carefully designed target layer that is placed over the sample being tested. With the high collection efficiency and resolving power of a TES spectrometer, we can isolate x-rays generated in the target from background and trace their paths through regions of interest in the sample layers, providing information about the various materials along the x-ray paths through their attenuation functions. We have recently demonstrated our approach using a 240 Mo/Cu bilayer TES prototype instrument on a simplified test sample containing features with sizes of $\sim 1$ μm. Currently, we are designing and building a 3000 Mo/Au bilayer TES spectrometer upgrade, which is expected to improve the imaging speed by factor of up to 60 through a combination of increased detector number and detector speed.

The work is supported by the Intelligence Advanced Research Projects Activity (IARPA) through the Rapid Analysis of Various Emerging Nanoelectronics (RAVEN) research program, agreements D2019-1906200003 and D2019-1908080004.

Sandia National Laboratories (Sandia) is a multimission laboratory managed and operated by National Technology & Engineering Solutions of Sandia, LLC, a wholly owned subsidiary of Honeywell International Inc., for the U.S. Department of Energy's National Nuclear Security Administration under contract DE-NA0003525. This paper describes objective technical results and analysis. Any subjective views or opinions that might be expressed in the paper do not necessarily represent the views of the U.S. Department of Energy or the United States Government.

P. Szypryt, D. A. Bennett, W. B. Doriese, M. Durkin, J. W. Fowler, E. J. Garboczi, G. C. Hilton, J. Imrek, V. Y. Kotsubo, Z. H. Levine, J. A. B. Mates, K. M. Morgan, N. Nakamura, G. C. O'Neil, N. J. Ortiz, C. G. Pappas, C. D. Reintsema, D. R. Schmidt, D. S. Swetz J. N. Ullom, and A. L. Wessels are with the National Institute of Standards and Technology, Boulder, CO 80305 USA (e-mail: paul.szypryt@nist.gov).

P. Szypryt, M. Durkin, J. F. Fowler, J. D. Gard, J. Imrek, K. M. Morgan, N. J. Ortiz, C. G. Pappas, J. C. Weber, and A. L. Wessels are with the Department of Physics, University of Colorado, Boulder, CO 80309 USA.

W. J. Boone, Amber L. Dagel, G. Dalton, E. S. Jimenez, K. Larson, D. McArthur, K. R. Thompson, C. Walker, and J. W. Wheeler are with Sandia National Laboratories, Albuquerque, NM 87185 USA.

## I. INTRODUCTION

INTEGRATED circuits (ICs) have become increasingly more complex, with tens of fabrication layers and feature sizes of $\lesssim 10$ nm. Although fabrication of circuits with decreasing feature sizes has been steadily progressing, methods for three-dimensionally imaging the many layers of such circuits post fabrication have lagged behind. This can be important for process development and defect analysis applications. Three-dimensional imaging can also be useful for microcircuit security and verification that externally sourced circuits do not contain unknown/unwanted structures. This three-dimensional imaging problem is reviewed in more detail in [1].

Recent work has shown that three-dimensionally imaging ICs at the 10 nm resolution scale is possible with x-ray ptychography [2], [3], but this method requires a highly coherent and bright x-ray source typically only available at a synchrotron. We are instead developing a tabletop approach that combines a scanning electron microscope (SEM) with x-ray microcalorimeters (hereafter μcals) based on the superconducting transition-edge sensor (TES; see [4], [5]) to perform x-ray tomography [6]. While the IC loses functionality during the sample preparation process, this approach is also non-destructive in that the measurement can be repeated. Our prototype instrument uses an array of 240 Mo/Cu TESs, but we are currently developing a larger instrument containing a 3000-pixel array of faster Mo/Au TESs with the primary goal of increasing imaging speed. To read out the larger array of faster detectors, we plan to use high bandwidth microwave SQUID multiplexing.

Our approach originated with the Non-destructive Statistical Estimation of Nanoscale Structures and Electronics (NSENSE; see [7], [8]) concept, and an overview of the approach is shown in Fig. 1. To begin, a target layer of known composition and thickness is deposited on the sample being tested after the sample has been thinned down to its active layers. During measurement, the electron beam from a SEM is focused onto the target layer, producing a small x-ray-generating region defined by the electron beam spot size.

X-rays generated at the target can travel through the sample layers, and the x-ray intensity is attenuated depending on the local composition and thickness along a given x-ray path. A fraction of these x-rays are detected by the TES μcals, and their arrival times and energies are determined in post-processing. Due to the energy resolution of our detectors, we can distinguish signal x-rays generated in the target from background x-rays generated elsewhere, such as the spacer or the SEM chamber walls. The sample stage is translated or the electron beam is scanned across the sample to change the location of the spot size relative to the sample layers, allowing us to generate a two-dimensional map of the attenuation across a region of interest on the sample. The sample is then rotated and once again scanned, producing another attenuation map, but at a different projection. Multiple such projections are fed into a tomographic reconstruction algorithm, resulting in a three-dimensional image of the sample.

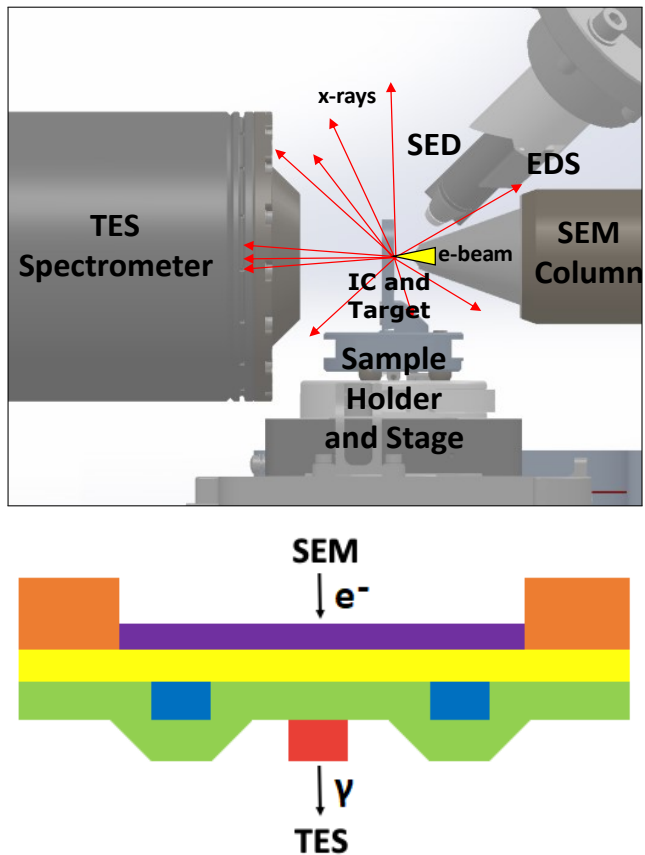

Fig. 1. (Top) layout of SEM chamber with prototype TES spectrometer and (bottom) example target/sample layer stack. During operation, the electron beam of a SEM is focused onto a target layer (purple), which generates x-rays in a localized spot. A spacer layer (yellow) is placed beneath the target layer, which is used to set the magnification and to block excess electrons from penetrating into the sample layers (blue, red, and green). X-rays generated in the target layer can be distinguished from background x-rays using the TES energy resolution, and these target x-rays are used to form attenuation maps of the IC. Other detectors shown in the top image provide supplemental information during a sample scan, including a secondary electron detector (SED) which can be used to map target surface topography and an energy-dispersive x-ray spectroscopy (EDS) detector which can be used to monitor x-ray generation variations in the target. In the test sample measurements described in Sec. II, the sample stack contained a 275 μm thick Si substrate (orange), a 70 nm thick Ti blanket target layer (purple), and a 2 μm thick $Si_3N_4$ membrane used as a spacer layer (yellow). The sample layers were composed of two 500 nm thick Nb features (blue and red) of varying lateral size with a 700 nm thick $SiO_2$ nonplanarized blanket layer (green) separating the two Nb layers.

## II. PROTOTYPE SPECTROMETER AND MEASUREMENTS

The current x-ray spectrometer is built into a two-stage adiabatic demagnetization refrigerator (ADR) with a detector protrusion that reduces the distance between the detectors and the x-ray source (similar design as NSLS cryostat described in [9]). The detectors are read out using time-division multiplexing (TDM [10]) with a two-stage superconducting quantum interference device (SQUID [11]) architecture [12]. The spectrometer consists of 240 μcals designed for x-ray energies of $\lesssim$14 keV. Unlike in many traditional TES applications where photon count rates on order 1 count per second (cps) per detector are expected, these μcals were optimized for performance at relatively high count rates ($\gtrsim$100 cps per detector). The energy resolution of these detectors is 12 eV at 8.0 keV (Cu K$\alpha$) and 200 cps per detector, as measured with a fit to a detector-coadded energy spectrum. A more detailed description of this spectrometer and associated performance measurements are presented in [13].

We fabricated a test sample, described in Fig. 1 (bottom), to use in demonstration measurements with the prototype spectrometer. We focused the measurement on an 30 μm × 3 μm area of the sample which contains 3 parallel Nb wires on 2 different layers separated by a 700 nm thick $SiO_2$ layer. The Nb wires are 500 nm thick and 2 μm wide, with a center to center lateral separation of 4 μm between wires. The fine motion stage was used to perform a serpentine scan across the sample with 500 nm steps, resulting in a total of 61 × 7 = 427 dwell locations. The stage step direction was at a ∼45° angle relative to the Nb wires. The electron beam energy was set to 15 keV and current was set to ∼75 nA. This resulted in roughly 100 cps per detector when the target was at normal incidence to the beam. A map of the fitted Ti K$\alpha$ integrated amplitude by dwell location was saved for each of the 9 projections (0°, ±15°, ±20°, ±25°, ±30° rotation angles) used in this initial scan.

The scan data at the 9 projections were fed into two independent tomographic reconstruction algorithms to produce a three-dimensional image of the sample. One of the attempted methods was based on a maximum likelihood estimation weighted with a maximum *a posteriori* (MAP) distribution [14], whereas the other method was based on the algebraic reconstruction technique (ART; see [15]). Both methods were optimized for the material inspection problem and adapted to handle the unique geometry of our system. They allow for irregular grid spacing and reconstruction with limited projection angles. These techniques both arrived at similar results, getting accurate values for the lateral separation of the Nb wires (∼200 nm lateral resolution), but both underestimated the interlayer spacing between the two Nb layers (worse than 500 nm depth resolution). In addition, both methods underestimated the density of the Nb wires as the measured attenuation contrast between regions with and without Nb was lower than the expected value. The cause of these issues is an ongoing research effort, but the leading explanation is a larger than expected electron beam spot size (>100 nm diameter). An example of a three-dimensional image of the test sample is shown in Fig. 2.

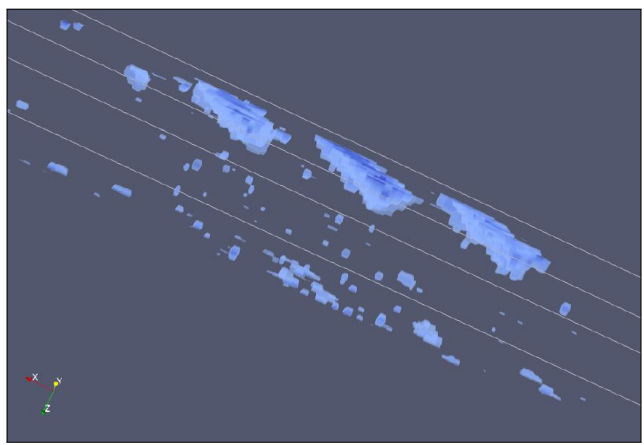

Fig. 2. Three-dimensional image created with maximum *a posteriori* (MAP) estimation technique. The code was built off of work presented in [16], which was originally developed to perform tomographic reconstructions including scatter corrections. Because the solid angle of the μcal array was small in our case, these scatter corrections could be ignored.

## III. 3000-PIXEL SPECTROMETER DEVELOPMENT

Imaging large areas, on order square millimeters, of an IC at small resolution scales (10–100 nm) takes a considerable amount of x-ray data, and the imaging speed depends largely on the x-ray collection capacity of the TES spectrometer and the SEM current. For this reason, we are building a larger TES spectrometer with 3000 total detectors to increase the active area of our system. This is an order of magnitude larger than the ∼250-pixel scale TES-based x-ray instruments commonly seen in the field. In addition, we are designing detectors with shorter response times to achieve increased per detector count rate limits. To accommodate these changes, the new system will use a dilution refrigerator in place of the current ADR and will use microwave SQUID multiplexing readout instead of TDM readout. Additionally, several data processing improvements have been developed to more efficiently analyze the large amount of x-ray data.

### A. Mechanical design and integration with SEM

The new spectrometer is being designed to be compatible with the existing SEM. The planned integration with the SEM is illustrated in Fig. 3. The spectrometer cryostat has a separate vacuum space from the SEM in order to keep the systems mechanically isolated and avoid coupling of vibrations from the cryostat to the SEM, which could result in image resolution degradation. The detector protrusion of the spectrometer is inserted into a receiver tube to reduce the distance between the detectors and the sample. This distance is primarily limited by the range of motion of the coarse motion hexapod stage, requiring a minimum separation between the detectors and sample of ∼120 mm. The SEM port size also limits the diameter of the detector protrusion to <250 mm, which ultimately limits the amount of area that could be used for the detectors. It should be noted that although minimizing the distance between the detectors and sample is important for maximizing flux, it may not necessarily be the best design choice for optimizing spatial resolution. For example, moving the detectors further back (or increasing detector density) could improve the theoretical limit on the achievable reconstructed spatial resolution. Additionally, placing detector elements away from the electron beam axis would increase the range of simultaneous projections that could be measured at a given sample rotation.

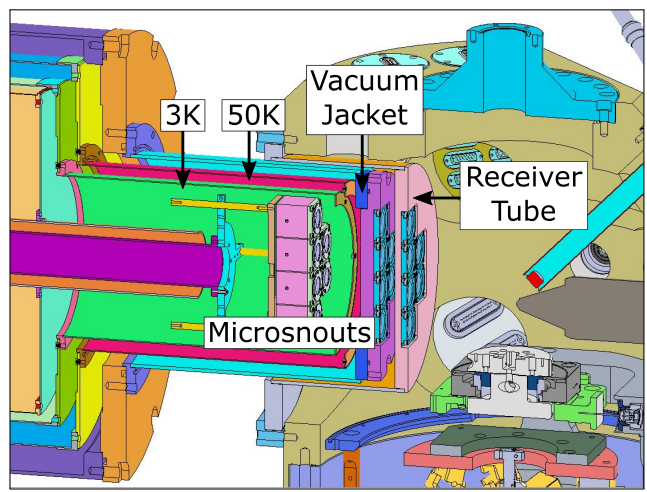

Fig. 3. Cross-section view of the 3000-TES spectrometer integrated with the SEM. The detectors are split into 12 "microsnouts" [17] (6 pictured in this cross-section, colored pink), or 250 detectors per microsnout, which are mounted to the cold stage (down to ∼10 mK) of the dilution refrigerator. These are enclosed by a set of shields at 3 K (green) and 50 K (red) that prevent 300 K radiation from reaching the detectors while providing magnetic shielding. Next, a vacuum jacket cylinder (blue) is placed over the 50 K cylinder. This is inserted into a receiver tube with diameter ∼250 mm, which is mounted to the SEM.

In order to allow x-rays to reach the detectors while blocking infrared, a set of thin Al filters are mounted to the fronts of the microsnouts as well as the fronts of the 3 K and 50 K radiation cylinders. The filters at the front of the microsnouts and 3 K shell are ∼100 nm thick, whereas the filters at 50 K are ∼200 nm thick to conduct the additional heat load expected at this stage. The vacuum jacket that is placed over the 50 K shell and the receiver tube mounted to the SEM also contains filters/windows through which x-rays can be transmitted. In addition, these windows are engineered to hold atmospheric pressure in order to maintain the two separate vacuum spaces. The windows on these components are designed to be ∼20 μm thick Be. The exact geometry and support structures for these windows are still being designed, but will need to accommodate the large angle x-ray paths expected with this system. Compared with the prototype system, the filter efficiency of this system is roughly 2× better in the 4–10 keV region of interest, or >80 % efficiency for x-ray energies >4 keV. This improvement is largely due to the overly cautious design of the prototype system which included 200 μm thick Be windows and a 5 μm thick Al electron filter between the receiver tube and vacuum jacket, which we have since found to be unnecessary.

### B. Microwave SQUID multiplexing readout

Microwave SQUID multiplexing, or μMUX, readout will be used in the new 3000-pixel instrument. A general de-scription of our implementation of this readout system is presented in [18], [19]; here we discuss some of the designed values specific for this spectrometer. The system will use 6 parallel readout channels, each with 4 GHz of bandwidth. Each readout line contains its own high electron mobility transistor (HEMT) amplifier, room temperature preamplifier, room temperature analog microwave electronics (IQ mixer, local oscillators, etc.), and digital electronics board including a field programmable gate array (FPGA). A total of 500 TES μcals (2 microsnouts) will be read out with each readout line. Each detector is coupled to a RF-SQUID and superconducting LC resonator with a designed bandwidth of 1 MHz. The designed frequency separation between resonators is 8 MHz. Due to the high photon rates expected with this system, particular care is being taken to mitigate cross-talk effects in the SQUIDs, resonators, and broadband amplifiers [20].

### C. Fast transition-edge sensors

The TES x-ray μcals for the 3000-pixel spectrometer are currently under development. They will consist of Mo/Au bilayer TESs with gold "sidecar" absorbers [21] coated in 15–20 μm of electroplated bismuth. The total area of each pixel will be the maximum that can fit on a 250-pixel microsnout array, about 500 μm × 500 μm. The saturation energy target is 30 keV, the maximum electron energy of our SEM column, and therefore the maximum x-ray energy that can be produced. We hope to achieve a multiplexed, high count-rate energy resolution of 10 eV or better to ensure minimal integration of background Poisson noise into our fluorescence signal lines. The detectors will be designed to be as fast as the 1 MHz μMUX readout can handle without significantly degrading the intrinsic detector energy resolution. The target pulse time, defined as the time for a pulse to decay to 10 % of its peak value, is 150–200 μs (700–800 μs in prototype instrument). Modelling suggests they could handle up to 1000 cps photon rates (200 cps in prototype instrument) with a low fraction of pulse pileup rejection ($\lesssim$30 %).

### D. Data acquisition and processing

The raw data signals measured by our spectrometer are pulses of current that represent the TES response to absorbed x-rays, transformed into more arbitrary analog-to-digital converter (ADC) units after being amplified by the set of SQUID and other amplifiers. Typically, the current pulses are recorded to disk and analyzed using the Microcalorimeter Analysis Software System (MASS) Python package [22]. This analysis package starts with raw pulse records, applies a set of cuts to remove piled up pulses, creates a set of optimal filters [23] for each detector that is insensitive to the sub-sample arrival time of the photon, and aligns the optimally filtered value spectra for each detector to that of a reference detector. Next, the filtered values are corrected for potential linear drifts and then mapped to physical energy values by fitting a cubic spline curve through the corrected filtered value spectra using known spectral features. The fitted line intensities of characteristic lines associated with the target material(s) separated by dwell location are the outputs that are then fed into the tomographic reconstruction algorithm.

Because we are working with count rates that are much higher than those of our typical TES applications, we have decided to offload much of the data processing to the data acquisition, reducing the amount of data that needs to be written to disk and the subsequent processing time. Prior to acquiring data on a sample of interest, we take pulse and noise training data that are used to create a set of projectors for each detector, similar to optimal filter creation. When x-rays are detected by the readout system, the software determines a filtered value for each x-ray with an optimal linear combination of the pulse projections, and these filtered values are written to disk instead of the full pulse records. We then proceed with the remaining analysis steps, as outlined above. We find no evidence for energy resolution degradation using this process and the calculated projectors do not change significantly throughout the course of a typical day-long measurement series.

## IV. CONCLUSIONS AND FUTURE WORK

The 3000-pixel instrument utilizes a dilution refrigerator and μMUX readout to achieve its high pixel counts and rates. We are expecting a 12.5 increase in x-ray collection capacity due to the increase in pixel number (3000 as opposed to 240) and another factor of ∼4–5 due to increased detector speed (1000 cps as opposed to 200 cps), for a total factor of ∼60 potential improvement over the prototype system. In order to utilize the improvement in detector speed, the absorber area and thickness will be larger, thinner filters with increased x-ray transmittance will be used, and SEM columns with increased flux capabilities will be investigated.

In addition to upgrading the TES spectrometer, we are also looking into other subsystems for potential improvement. For example, we plan to use nanoscale structured targets patterned with electron beam lithography to achieve image spatial resolution comparable to the size of these structures, smaller than the intrinsic electron beam spot size. These nanostructured targets can be made arbitrarily complex with multiple materials that provide characteristic x-ray line markers across a wide range of energies, providing additional information about the energy dependent attenuation functions of materials within a sample being tested. We are also looking into resolving excess vibration and drift issues in the current SEM and are also looking into a potential SEM column upgrade aimed at increasing the overall electron flux density.

We are also looking into potential spectrometer upgrades beyond the 3000-pixel design, aimed at further improving imaging speed and resolution. This can be done in several ways, the simplest of which would be to increase the SEM chamber size and build up additional microsnouts and associated readout channels. Alternatively, we are developing a "nanosnout" assembly technology that replaces macroscopic microwave cables with compact flexible cables and aluminum wire bonds with indium bump bonds, both of which are used to increase the pixel packing density. Finally, we are looking at other detector technologies, such as x-ray μcals based on kinetic inductance detectors (KIDs; see [24], [25]), to replace the TES-based µcals. Although the energy resolution performance of this detector technology is far behind that of TESs, it has the potential for denser detector arrays and much simpler scalability of the readout. All in all, there is a large parameter space for developing a x-ray tomography system using low-temperature, energy-resolving detectors, and the ideal system configuration for a future instrument is yet to be determined.

ACKNOWLEDGMENT

The authors would like to thank Eugene Lavely, Adam Marcinuk, Paul Moffitt, Steve O'Neill, Thomas Stark, Chris Willis and others at BAE Systems for their role in the NSENSE concept development and initial instrument integration in the NIST Boulder laboratories.